\documentclass[preprint]{aastex}

\begin{document}

\title{$JHK$ Magnitudes for L and T Dwarfs and Infrared Photometric Systems}

\author{D.C. Stephens}
\affil{Space Telescope Science Institute, Baltimore, MD 21218}
\email{stephens@stsci.edu}

\and

\author{S.K. Leggett}
\affil{Joint Astronomy Centre, University Park, Hilo, HI 96720}
\email{s.leggett@jach.hawaii.edu}

\begin{abstract}
In the last few years a significant population of ultracool L and T dwarfs 
has been discovered.  With effective temperatures ranging from $\sim$ 2200 to 700 K, these 
objects emit most of their radiation in the near infrared and their spectral 
energy distributions are dominated by strong molecular absorption bands.
These highly structured energy distributions lead to $JHK$ 
magnitudes that are extremely sensitive to the exact filter bandpass used.  In the 
case of the T dwarfs, the differences between commonly used photometric systems can be 
as large as 0.4 mag at $J$ and 0.5 mag at $J-K$.  

Near--infrared magnitudes have been published for L and T dwarfs
using a variety of photometric systems.  Currently, the data obtained with these
systems cannot be accurately compared or combined as transformations based on the
colors of hotter
stars are not valid for L and T dwarfs.  To address this problem, we 
have synthesized $J, H$, and $K$ magnitudes for some of the common photometric 
systems and present transformation equations with respect to the most 
atmospheric--independent system, the Mauna Kea Observatory (MKO) filter set.  If the 
spectral type of the dwarf is known, our transformations allow data to be 
converted between systems to 0.01 mag, which is better than the typical 
measurement uncertainty.  Transforming on the basis of color alone is more difficult 
because of the degeneracy and intrinsic scatter in the near--infrared colors of 
L and T dwarfs; in this case $J$ magnitudes can only be transformed to $\lesssim$0.05 
mag and $H$ and $K$ to $\lesssim$0.02 mag.

\end{abstract}


\section{Introduction}

L and T dwarfs have unusual spectral energy distributions (SEDs), with most of their 
flux emitted through windows in the near--infrared.  Normalized spectra of 
an L5 and T4.5 dwarf \citep{geb02} are shown in Figure 1 to
illustrate how absorption bands of $\rm H_{2}O$, CO, and $\rm CH_{4}$ 
regulate the near--infrared emission and create the flux windows.
The $\rm H_{2}O$ bands 
are the same features responsible for the telluric absorption that 
defines the conventional $J$ (1.1---1.4 $\micron$), $H$ (1.5---1.8 $\micron$), and $K$ 
(2.0---2.4 $\micron$) bandpasses.  Thus, the presence of $\rm H_{2}O$ in the atmospheres 
of the L and T dwarfs forces much of their flux to be emitted within these bands, resulting 
in the extreme far--optical and near--infrared colors that are used to identify L and T dwarf 
candidates from photometric surveys like the 2 Micron All Sky Survey (2MASS, \citealt{2m}), 
and the Sloan Digital Sky Survey (SDSS, \citealt{sdss}).  

There are now hundreds of L and T dwarfs with near--infrared magnitudes published in 
various photometric systems.  To maximize the science potential of these observations and 
their impact on brown dwarf theory, transformation equations are desirable to convert these 
and any future magnitudes to a common photometric system.  Because the $J$, $H$ and $K$ 
bandpasses include  $\rm H_{2}O$, CO, and $\rm CH_{4}$ absorption features, any variation in 
the width of the filters 
will lead to system--dependent magnitudes.  Therefore, transformation equations need to be 
derived as a function of near--infrared spectral type or, less desirably, color (see later 
discussion in \S 4) to correctly account for the presence of the molecular bands.  

Figure 2 shows $J-K$ color as a function of spectral type for typical main sequence stars 
(A0--M5) in the Bruzual--Persson--Gunn--Stryker 
Atlas\footnote{http://www.stsci.edu/instruments/observatory/cdbs/astronomical\_catalogs.html}, 
and late M, L and T dwarfs reported in \citet{le02}.  Important 
features to note, all of which are explained more fully in other works (e.g. \citealt{le02}) are:
the intrinsic spread in $J-K$ for the L dwarfs that may be produced by variations in the extent 
and location of the condensed grain layer in the photosphere; the increasingly bluer $J-K$ color 
for the late L dwarfs and T dwarfs that is mostly due to the appearance of the $\rm CH_{4}$ band 
at 2.2\micron; and the scatter in $J-K$ for the latest T dwarfs likely due to 
gravity--dependent $\rm H_{2}$ opacity.  Although T dwarfs can have the same value of $J-K$ as 
A through M stars, convolving a T dwarf spectrum with its strong molecular absorption bands
(Figure 1) with any $JHK$ bandpass will produce a very
different result from the convolution of e.g. the Rayleigh Jeans curve of an A0 star with
the bandpass. Consequently, transformations based on the colors of hotter
stars are not valid for L and T dwarfs and
photometric transformations must be derived directly from observations
of these ultracool dwarfs.

In this paper we present synthesized $J, H$ and $K$ magnitudes
using near--infrared spectroscopic observations of L and T dwarfs from each spectral subtype,
for the photometric systems in which L and T dwarf photometry has most frequently been 
published and for established systems in which future observations may occur.  The 
photometric systems presented are: 
2MASS \citep{c01}, Caltech  (CIT, \citealt{el82}), the DEep Near-Infrared Survey (DENIS, 
\citealt{fou00}), Las Campanas Observatory (LCO, \citealt{per98}), Mauna Kea Observatory 
(MKO, \citealt{sim02,tok02}), the United States Naval Observatory Flagstaff Station (NOFS, 
\citealt{d02,gue03}) 
and the United Kingdom Infrared Telescope (UKIRT, \citealt{haw01}).  We 
also generate equations  to transform $J, H$ and $K$ magnitudes 
between the other systems and the MKO system. The MKO photometric system 
was chosen as the reference point because MKO filters are narrower than classical 
$J$, $H$ and $K$ filters, thus avoiding the telluric absorption bands that can vary with time 
and observing location (see discussion in \citealt{sim02}, \citealt{tok02}).  As a result
MKO magnitudes have little dependence on local observing conditions, and  their use produces 
transformation equations with less uncertainty than would be obtained using another photometric 
system.  In addition, the MKO filters have been widely adopted and the system is endorsed by 
the IAU Working Group on Infrared 
Photometry as the preferred photometric system for ground-based near--infrared observations.  

In \S 2 we present the observed differences in the 2MASS and MKO magnitudes 
that have been measured for several L and T dwarfs, showing that system transformations 
cannot be reliably derived empirically due to significant uncertainty in the observational data.  
\S 3 discusses the inputs for synthesizing magnitudes: filter transmission profiles, telluric
absorption bands, instrument optics and observed spectra.
Our results are presented in \S 4 and our conclusions given in \S 5.

\section{Observed Magnitudes in Different Systems}

Figures 3 and 4 compare the $J, H$ and $K$ magnitudes, and $J-H, H-K$ and $J-K$ colors
for a sample of L and T dwarfs that have been observed in both the 2MASS and MKO
photometric systems.  These are the only systems with a large
enough number of L and T dwarfs in common to make a meaningful observational comparison.
2MASS magnitudes for these objects were taken from the 2MASS L dwarf archive 
webpage\footnote{$\rm http://spider.ipac.caltech.edu/staff/davy/ARCHIVE/index\_l\_spec.html$} 
and A. Burgasser's T dwarf 
webpage\footnote{$\rm http://www.astro.ucla.edu/\sim adam/homepage/research/tdwarf/$}.
The MKO magnitudes are reported in \citet{le02}.
Figure 3 plots $\delta$mag as a function of spectral type and Figure 4 plots
$\delta$mag as a function of $J-K$ (on the MKO system), which provides the largest
baseline.

Spectral type is taken from \citet{geb02}, who define a classification 
scheme for both L and T dwarfs based on the strength of the near--infrared 
absorption bands.  This classification gives results very similar to the 
scheme presented for the T dwarfs by \citet{burg02}.  However, 
the scheme for L dwarfs presented by \citet{k00}, which is 
based on red spectra, can assign L dwarf spectral types that differ by up to 
2.5 subclasses from the Geballe classification.  For the samples shown in Figures 
3 and 4 the average difference in L dwarf classification is only 1.0 subclass.
Therefore, given the size of the observational uncertainty (see the Figures) the 
choice of classification scheme is not significant.

Despite the large observational uncertainty in Figures 3 and 4, it can be seen
that there are significant differences in the magnitudes, especially at $J$,
and that general trends in $\delta$mag with type do exist.  The difference between
systems can be
understood with reference to the spectra shown in Figure 1.  The 2MASS $J$ 
filter is wider than the MKO $J$ and 2MASS $K$ is narrower than MKO $K$ 
(\S 3.1);  the wider filters include more of the absorption bands of 
$\rm H_{2}O$ and $\rm CH_{4}$ without
increasing the signal, and hence, with reference to the calibrator, L and T 
dwarfs appear to be fainter in the systems with wider filters.  
As these features become stronger with later spectral type, the effect is more
pronounced and trends appear with 2MASS $J$ becoming increasingly fainter than MKO 
$J$, and MKO $K$ fainter than 2MASS $K$.  Although these trends can be seen, 
the considerable uncertainty in the data prevents 
the determination of reliable system transformations from direct observations.  
In the following sections we derive and discuss theoretical
transformations between these and other systems.

\section{Calculation of Synthetic Magnitudes}

\subsection{Filters}

Figure 5 shows the filter profiles for the 2MASS, CIT, DENIS, LCO, MKO, NOFS and UKIRT  
$JHK$ filters at instrument temperatures.  The 2MASS filter profiles were obtained 
from the 2MASS 
webpages\footnote{$\rm http://www.ipac.caltech.edu/2mass/releases/second/doc/sec3\_1b1.html$}, 
the MKO filter profiles were obtained from the UKIRT 
webpages\footnote{$\rm http://www.jach.hawaii.edu/JACpublic/UKIRT/instruments/uist/imaging/filters.html$}
and the UKIRT profiles from \cite{haw01}.  
The LCO profiles were generated with tables obtained from 
\citet{per98}, the DENIS profiles were obtained from P. Fouqu\'{e} 
(priv. comm. 2002) and the NOFS profiles were obtained from F. Vrba (priv. comm. 2003).
The CIT--$H$ and $K$ profiles measured at operating temperature
were obtained from the CTIO infrared instrumentation 
webpage\footnote{$\rm http://www.ctio.noao.edu/instruments/ir\_instruments/irfilters/filters.html$}, 
where they are identified 
as 25mm OCLI $H$ and $K$ filters.  We selected these filters for CIT--$H$ and $K$ because 
they match the documented CIT bandpasses \citep{el82}.  The CIT--$J$ profile 
measured at ambient temperature was also obtained from the CTIO webpage, where it is
identified as  CIT--$J$.  A shift to bluer wavelengths was required to correct this 
profile to values appropriate for operating temperatures.  

We attempted to determine the appropriate shift for the CIT--$J$ band from the literature,
however there is a discrepancy between the cold transmission profile measured for 
CIT--$J$ by H. Jones (private comm. between E. Persson and H. Jones 1994; \citealt{jon94}) 
and the bandpass given by \citet{el82}.  Therefore, two independent shifts were made to the 
CIT--$J$ bandpass, producing two different transmission profiles.  The first 
shift moved the bandpass $\sim$0.015 $\mu$m and was chosen 
to produce a bandpass identical to the one determined for the CIT system by H. Jones.  We 
refer to this filter bandpass as CIT--$J$ throughout the rest of the paper (solid line
in Figure 5).  The second $J$ bandpass was created by shifting the ambient $J$ profile 
$\sim$0.04 $\mu$m to match the CIT--$J$ bandpass specified by \citet{el82}.  In determining 
this shift, we assume that the \citet{el82} bandpass limits do not include atmosphere (see \S 3.2) 
and we refer to this bandpass in the paper as Elias--$J$ (dash--dot line in Figure 5).

Since the original 1980--era CIT filter profiles no longer exist (J. Elias and E. Persson,
private comm. 2003), it is not clear which transmission profile represents the original 
CIT--$J$ filter.  The Elias--$J$ bandpass involves a $\sim$3\% shift in wavelength of the 
webpage ambient profile, which is about twice the value seen for the UKIRT filters on a 
cooldown from ambient to 77K.  In this regard, the bandpass measured by Jones seems more
reasonable.  However, we present both profiles in this work as the bluer bandpass 
is specified in the defining work by \citet{el82}.  Note that the red cut-off of 
each filter is effectively defined by the atmosphere, but the differences in the 
blue cut--on produces significant systematic differences in the $J$ magnitudes, as we show in \S 4.  
The $H$ and $K$ magnitudes are well determined as these profiles agree with the \citet{el82} 
bandpasses and are identical to the CIT profiles measured by H. Jones 
(private comm. 1994; \citealt{jon94}).

\subsection{Atmospheres}

Figure 5 shows as a dotted line the effective bandpass of each filter after convolving
with the atmospheric transmission appropriate for each site.  
For 2MASS and NOFS the mean transmission
of Mount Hopkins (which appears to be equivalent to a little more than 5mm water) was used 
\citep{c01}.  For CIT, Elias--$J$ (dashed line), DENIS, and LCO an atmosphere profile 
typical of Las Campanas was obtained from the Las Campanas WIRC Users Manual 
webpage\footnote{$\rm http://www.ociw.edu/instrumentation/wirc/wirc.html$}. For UKIRT and 
MKO the 1.2~mm Mauna Kea atmosphere was used.  The Mauna Kea atmosphere for various 
values of water vapor have been calculated by the ATRAN model \citep{lo92} and are 
available from the Gemini 
webpages\footnote{$\rm http://www.gemini.edu/sciops/ObsProcess/obsConstraints/ocTransSpectra.html$}.  
An atmosphere profile typical of conditions 
at Cerro Tololo was also obtained that could have been used with the CIT filters instead
of the LCO atmosphere.  However, the difference in synthetic CIT magnitudes produced 
from the two atmospheres was negligible for $H$ and $K$, and never more than 
0.005~mag at $J$ for the L dwarfs and 0.009~mag at $J$ for the T dwarfs.  

Figure 5 shows that all but the LCO and MKO $J$ filter bandpasses extend into poor regions 
of the atmosphere, usually by being too red (although the 2MASS, DENIS and NOFS filters are 
also too blue).  \citet{c01} states that the 2MASS $J$--band calibration zero points often 
showed variations within a night as large as 0.1~mag,  most likely due to variations in the 
atmosphere (however colors were stable to $<$0.02~mag).  To explore the effect of variable 
water vapor on the $J$ magnitude, we synthesized photometry for each system using 
different amounts of water vapor.  We found that if the water vapor is varied between 
1~mm and 3~mm, the UKIRT $J$ magnitudes differ by 0.01~mag for mid--L dwarfs, 0.02~mag for 
late--Ls, 0.03~mag for early--Ts, and 0.05~mag for late--T dwarfs.  If the water vapor is 
varied between 3~mm and the mean transmission at Mount Hopkins ($\sim$5mm), the 2MASS $J$ 
magnitudes differ by 0.03~mag for the early-- to mid--L dwarfs, 
0.04~mag for the late--Ls, and from 0.05~mag up to 0.1~mag for the T dwarfs.  The dependence
of the $J$ magnitude on the atmospheric transmission highlights the need for a bandpass 
that is free of the atmosphere (see also further discussion in \citealt{sim02}, \citealt{tok02}).

\subsection{Detector and Optical Responses}

The filter profiles shown in Figure 5 do not include the effect of
other optical elements in the instrument lightpath.  Telescope mirrors, instrument
optics and the detector quantum efficiency (QE) will produce wavelength--dependent
transmission or reflection curves which should, strictly, be convolved with the 
filter and atmosphere transmission.  We have investigated the effect of various
commonly used elements and show below that they are negligible;  therefore, 
we calculate synthetic magnitudes using the  filter and atmosphere transmissions
only.

The commonly used reflective surfaces in a near--infrared telescope will be gold,
silver or aluminum.  The reflection curves of these surfaces are flat within
measurement uncertainty from 1.0 to 2.6 $\mu$m, at 94 to 98\%.  The commonly used
transmissive elements are zinc selenide, and calcium, lithium or barium fluoride.
Uncoated, such windows or lenses have flat transmissions at 94 to 95\%.
Anti--reflection (AR) coatings can lead to a wavelength--dependent response, however
the coatings for infrared instrument are usually optimized for the near--infrared
and the lens coatings in the UKIRT cameras for example vary in throughput by only 
1\%, typically,  across any of the $J$, $H$ or $K$ bandpasses.

A more serious issue is variations in detector QE, in particular large variations
in the AR coatings on detectors.  We have found that 
current 1024$\times$1024 InSb ALADDIN arrays have a more structured reflectivity
curve than older generation InSb detectors.  UKIRT's ALADDIN detector in the UIST 
camera has a curve that varies in reflectivity from 10 to 20\% across the $J$ filter
for example, while the older 256$\times$256 InSb detector in IRCAM is reasonably
flat with a $\leq$2\% change across any filter bandpass.  \citet{c01} 
indicates that the 2MASS NICMOS detector QE is quite flat at 60 to 65\% across
$J$, $H$ and $K$.
Given the highly structured SEDs of the L and T dwarfs we investigated the impact
of the detector coatings by calculating synthetic magnitudes for the MKO filter set 
with and without the ALADDIN--type coating.  We find the effect to be 0.01 mag at
$J$ for mid--L through T types, 0.002 mag at $H$, and 0.003 mag at $K$ except
for late T dwarfs where the difference is 0.01 mag at $K$.  Measurement uncertainties 
are always significantly larger than these values, hence it appears that AR coatings,
while important for instrument throughput, have a negligible effect on
photometric systems, even for L and T dwarfs.

\subsection{Spectra}

Magnitudes were synthesized by convolving observed flux--calibrated spectra across 
the filter$+$atmosphere bandpasses, and calibrated by 
convolving these same bandpasses with the observed energy distribution of Vega 
\citep{hay85, mnt85} which was assumed to have zero 
magnitude at each bandpass.  The L and T dwarf spectra 
were taken from \citet{geb02}, supplemented by additional data on 
2MASS0415$-$09 and SDSS1110$+$01 \citep{kn04}.  
The noise in the L and T dwarf spectra was typically
2\% of the flux at $J$ (rising to 4\% for the $\sim$30\% of the sample with $J\approx 17$),
1\% at $H$ (rising to 2\% for the faintest dwarfs), and 3\% at $K$ (rising to 4\%).
The uncertainty in the absolute flux calibration of Vega is estimated to be 3\% in the 
near--infrared.  However the uncertainty in the Vega calibration and in the input spectra 
can be neglected here.  The uncertainty in the
the flux calibration of the L and T dwarf spectra will be predominantly
due to the uncertainty in the photometry used to originally calibrate the spectra, 
which is 3--5\% 
\citep{le02}\footnote{The observed magnitude $m_o$ measured with a filter$+$atmosphere
profile $t_o$ is used to scale the input spectrum $f_*$ by a constant $c$ such that
$$ 10^{-0.4m_o} = c*\case{\int f_*t_o\delta\lambda}{\int f_{Vega}t_o\delta\lambda} $$
then the derived magnitude $m^{\prime}$ is determined by
$$  10^{-0.4m^{\prime}} = 10^{-0.4m_o}*\case{\int f_*t^{\prime}\delta\lambda}{\int f_*t_o\delta\lambda}
* \case{\int f_{Vega}t_o\delta\lambda}{\int f_{Vega}t^{\prime}\delta\lambda} $$
While the absolute flux of Vega is uncertain at the 3\% level, the slope
across the filter is well determined.  The noise in the dwarf spectrum is only
significant where it is differently weighted by the two filter profiles.
Hence the uncertainty in the synthetic magnitude is driven by the uncertainty in $m_o$.}.

We found when deriving the synthetic magnitudes that more consistent results were 
obtained if the filter bandpasses all had the same, high resolution.  Therefore we 
resampled each filter profile to $\delta \lambda=$0.001~$\mu$m before convolving 
with the atmosphere.  The bandpasses used in this work, with and without the atmosphere 
(see \S 3.2), are available on request from the authors; they can be used to derive 
magnitudes in various
photometric systems from a flux--calibrated near--infrared spectrum, provided the
spectrum covers the full bandpass of the filter.

We also found that unless we 
had spectra that covered the complete bandpass, additional scatter was introduced 
into our magnitude comparisons.  This was only an issue for the $J$ filters as most
of the spectra did not go into the poor transmission region around 1.35~$\mu$m.  If 
incomplete spectra were used, errors were introduced in the synthetic 
magnitudes as large as 0.1---0.4 mag for the UKIRT filter, which has significant 
transmission at 1.35~$\mu$m.  For the other wide $J$--band filters the effect was smaller, 
typically 0.02 to 0.04 mag.  We corrected the spectra by interpolating the data across 
the gap using as templates bright dwarfs that had been observed in this spectral region 
on Mauna Kea.  Tests using a variety of templates show that the uncertainty introduced 
in the $J$ magnitude by this interpolation is $\lesssim$5 millimags for the wider filters, 
such as the UKIRT filter.

\section{Results}

Tables 1, 2 and 3 list the synthesized $J$, $H$ and $K$ magnitudes, respectively, 
in the various systems for the sample
of 24 L dwarfs and 17 T dwarfs. The magnitudes are given to the millimag level despite the
0.03--0.05 mag uncertainty in these derived magnitudes because we wish to avoid 
introducing errors in the transformations 
by rounding off the synthetic magnitudes to too low a level of significance.
The transformations are given by the difference in magnitudes from each system 
and as systematic errors in the flux calibration cancel out, 
the uncertainty in this difference is smaller than that in the
original photometry.  The uncertainty in the transformation is determined only by the 
weighting of the noise in the spectra by each bandpass, 
and also by any uncertainty in the definition of the bandpass (see \S 3.1, 3.2, 3.4 and 
further discussion below).

Figure 6 shows the calculated differences in $JHK$ for the various photometric systems
as a function of spectral type, and Figure 7 the difference in the colors
as a function of type.  Figures 8 and 9 also show $\delta$mag and $\delta$color,
this time as a function of $J-K_{MKO}$.  The trends in $\delta$(2MASS$-$MKO) agree 
well with the observed trends shown in 
Figures 3 and 4.  Other observational comparisons are very limited. 
Only three L dwarfs have independently measured $J$ and $K$ in both the
DENIS and MKO systems and for these the agreement with Figure 6 is reasonable, but the 
observational uncertainties are large.  No data currently exists in the LCO system for L and 
T dwarfs.  Data have been published for some dwarfs in both the UKIRT and MKO systems, but
these are not independent measurements, instead one dataset has been synthesized
from the other as we have done here.  Photometry for several L and T dwarfs obtained 
with the NOFS filters has been published by \citet{d02} and calibrated 
using \citet{el82} standards \citep{gue03}.  Comparison of the NOFS--system 
magnitudes with MKO data for 15 L dwarfs and two T dwarfs in common 
produces $\delta$mags which agree well with our derived sequences at $J$ and $H$, but
which differ at $K$ for the two T dwarfs by $\sim$0.2~mag (compared to the measurement
uncertainty of $\sim$0.1~mag).  These results will be investigated further
when more data from this group are available.

In Figures 6 and 7 spectral type is given on the infrared typing scheme of 
\citet{geb02}.  As discussed in \S 2, the optically--based L 
dwarf classification scheme of \citet{k00} can lead to differences in spectral
classification of up to 2.5 subclasses, although the average difference for this sample is 
only 1.0 subclass. The uncertainty in the Geballe classification is typically 0.5 
subclasses, and, given the slow change in $\delta$mag with type for L dwarfs, the 
sequences shown in Figures 6 and 7 should be effectively independent of the classification 
scheme.  We tested this by fitting $\delta$mag values using both the Geballe and 
Kirkpatrick classifications and the difference for a given L type was 
always substantially less than the standard deviation of the fit.  

As an additional check, we determined the difference in $\delta$mag that would 
occur if spectral type was allowed to vary by two L sub-classes, simulating the
difference in classification that can occur between the visible and 
near-infrared classification systems.  For the case of the 2MASS $J$ filter, 
which is a relatively steep function of type, the difference in $\delta$mag 
between L6 and L8 types is 0.017 mag.  The sensitivity to type for earlier L 
spectral types is $<$0.01 mag, and the $H$ and $K$ filters are insensitive 
to L dwarf type as can be seen in Figure 6.  Hence uncertainties in L 
spectral type lead to uncertainties in $\delta$mag of $<$0.01 mag except
for late L dwarfs at $J$ where large uncertainties could lead to an 
uncertainty of $\sim$0.02 mag.  Note that to determine near--infrared photometric 
system dependencies, an infrared scheme is more appropriate than an optical scheme
and should give a tighter relationship between $\delta$mag and type.

The $\delta$mag:type sequences can be fit well mathematically.  Table 4
gives the results of cubic fits to $\delta J$, $\delta H$ and $\delta K$
as a function of type, all with respect to the MKO system, and the standard
error of the fit in magnitudes (colors can be calculated by differencing the
relations).  The accuracy of the derived transformations are quite good --- the 
standard error is better than 0.01 mag.  The fits to $\delta$mag with 
type can be seen as solid lines in Figure 6.  The scatter around the fits 
is small, 0.005--0.020 magnitudes, which is consistent with the 
noise in the spectra.

The $\delta$mag:$J-K$ relationship is more difficult to fit due to the 
degeneracy in colors between early L and T dwarfs, the degeneracy within 
the L dwarfs, and the intrinsic spread in $JHK$ colors of L and T dwarfs 
with the same spectral type (\S 1, Figure 2).  
Objects with different spectral morphologies can have the same color, but will 
have different values for $\delta$mag.  Consequently, transformations based on 
color alone will combine dwarfs with different spectral characteristics and 
produce a $\delta$mag value which will be less accurate than the value based
on spectral type.  This is a problem in particular for $\delta J$, as a function
of $J-K$, as can be seen in Figure 8.  
Tables 5 and 6 give the results of fits to $\delta J$, 
$\delta H$ and $\delta K$ as a function of $J-K$ color in each of the 
photometric systems:  Table 5 gives the coefficients for the quadratic fit 
found for $\delta J$, and Table 6 gives the coefficients for $\delta H$ and $\delta K$
which were fit well with linear equations.  These fits are shown as solid lines in 
Figure 8.  The scatter around the fits at $J$ is 
0.005---0.070 mag, at $H$ it is 0.003---0.030 mag and at $K$  0.015---0.030 mag.  Thus 
$H$ and $K$ magnitudes can be transformed almost equally well using 
either spectral type or color, but $J$ magnitudes transformed from color 
will be much more uncertain than those based on type.  Separating the L and 
T dwarfs can produce better transformations for the $J$ filter; these fits, and 
fits using other color combinations, can be determined using the synthetic magnitudes
given in Tables 1---3.

To summarize, if the spectral type of the dwarf is known and the filter can be regarded 
as well determined, then $J$, $H$ and $K$ transformations can be determined to 
$\sim$ 0.01 mag using the equations provided in Table 4.  Therefore for most observations 
the original measurement uncertainty (typically $\gtrsim$ 0.03 mag, see e.g. Figure 3)
will limit the accuracy of the transformed magnitude . However
other uncertainties do exist that can significantly effect the $J$ magnitudes, or 
colors involving $J$. The CIT $J$ bandpass is not well known, and the profiles of
the wider $J$ filters are determined by a possibly variable atmosphere;
we showed in \S 3.2 that a plausible range in water vapor
levels  can lead to variations in the $J$ magnitudes of 0.05---0.10~mag for T dwarfs,
for such filters.  For late L dwarfs, an uncertainty in spectral type of
2 subclasses can lead to an uncertainty in  $\delta J \sim$ 0.02 mag.  If the spectral
type is not known at all, then transforming based on $J-K$ color leads to an
uncertainty in  $\delta J \sim$ 0.05 mag.  For mid--L through T types, variations
between detectors can result in an additional, but small, uncertainty in  
$\delta J \sim$ 0.01 mag (\S 3.3). 
The $H$ and $K$ bandpasses are better behaved.  Detector response uncertainty is only
expected to impact $K$ and then only for late T dwarfs at the 0.01 mag level.
Also, for $H$ and $K$ transformations can be derived to $\sim$ 0.02 mag
on the basis of color alone.

\section{Conclusions}

To obtain accurate and stable photometry, filter bandpasses should not go into 
poor regions of the terrestrial atmosphere  --- most (classical) $J$  filters 
are poorly defined from this point of view.  Variations in the water vapor content change 
the effective bandpass of such filters and, for objects with extremely structured spectral 
energy distributions  such as T dwarfs, these changes produce 
photometric deviations of  $\sim$0.05---0.10~magnitudes.  To measure magnitudes and colors
to better than this requires use of a filter set that is well matched to the 
atmosphere, such as the MKO filter set. 

$JHK$ magnitudes for L and T dwarfs are highly dependent on the photometric system used
for the observation; for T dwarfs differences between systems can be several tenths
of a  magnitude.  However, we have shown that $JHK$ magnitudes for L and T dwarfs can
be transformed between the 2MASS, CIT--$H$\&$K$, DENIS, LCO, NOFS and UKIRT systems and the 
MKO system to  $\sim$0.01~mag if the spectral type of the dwarf is known.
This is significantly better than the typical measurement uncertainty, i.e. the original
uncertainty in the measurement will determine the accuracy of the transformed value.  
For the CIT system the uncertainty in the $J$ bandpass effects the derived magnitudes 
by 0.05--0.10 mag for L and T dwarfs.
Variations between the optical elements of common infrared
instrumentation are expected to impact the measured magnitudes of the late L and
T dwarfs at the 0.01~mag level.
If spectral type is not known, then $J-K$ color can be used to
transform $H$ and $K$ magnitudes measured in different systems with an accuracy of about 
0.02~mag but the $J$ value can only be derived to $\sim$ 0.05~mag on the
basis of color alone.  

The results presented here will be valuable for researchers in the very active field of 
ultracool dwarf studies, where imaging data
are plentiful, and where the data have unfortunately been obtained with a variety
of photometric systems.  Transformations based on the colors of hotter stars, even if the
stars have
the same color, cannot be applied to objects with strong molecular absorption bands such as
those seen in L and T dwarfs.  For these ultracool objects 
intercomparison of photometric data requires knowledge of the filter profiles at instrument 
temperature, and knowledge of the local atmospheric transmission.  If $JHK$ photometry
is obtained with a well understood photometric system we have shown that such datasets
can be accurately combined or compared.

\acknowledgments

We are grateful to Jill Knapp and colleagues for use of spectra prepublication,
to Hugh Harris for useful discussions and to the referee Adam Burgasser for improvements
to the paper.
We thank all those responsible for setting up the 2MASS, DENIS and SDSS surveys,
which enabled the discovery of the population of field L and T dwarfs.

\newpage 

\begin{figure}
\includegraphics[scale=.65,angle=-90]{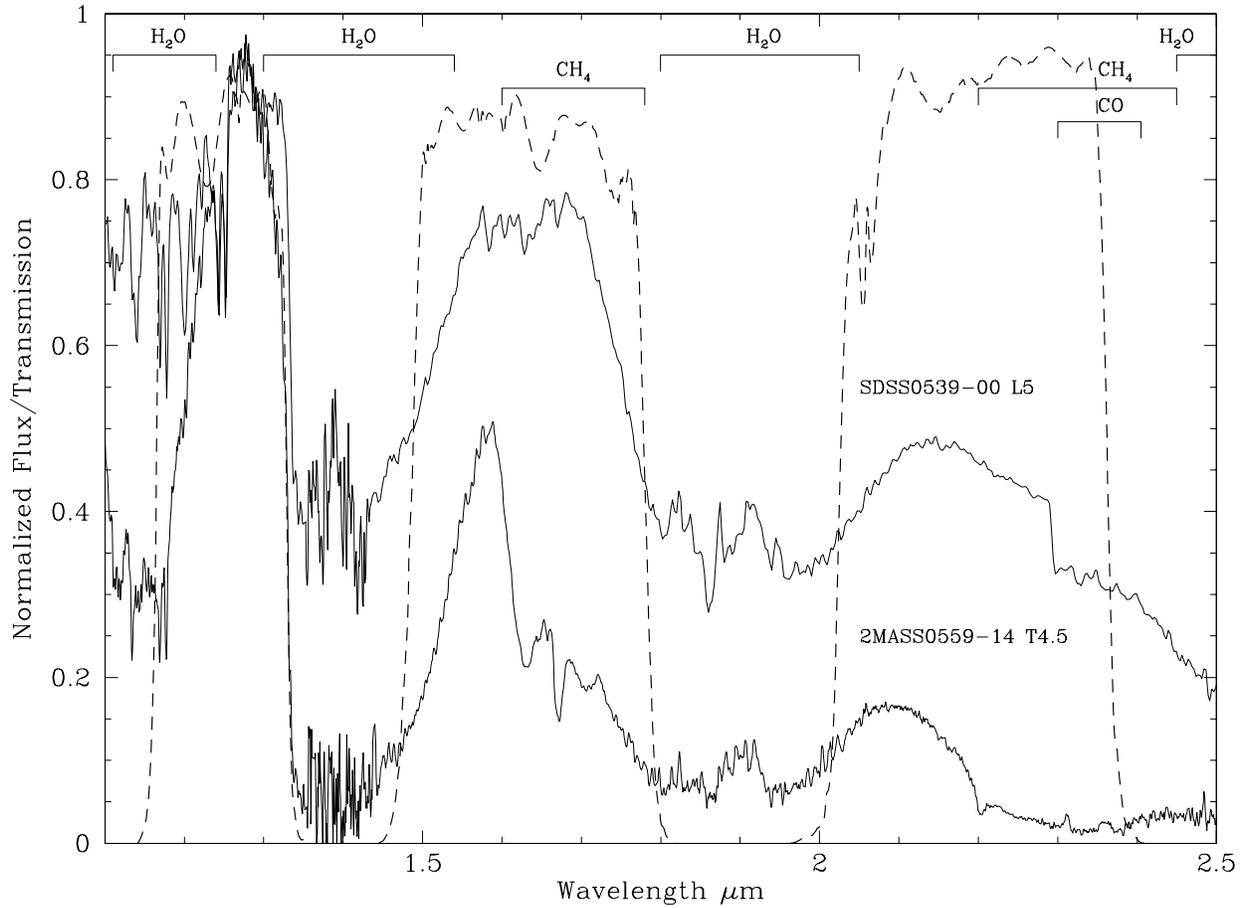}
\caption{Normalized observed spectra for an L5 and a T4.5 dwarf from \citet{geb02}.  The principal
absorption bands in the dwarf spectra are identified and the bandpasses for the MKO $J, H$ and $K$
filters (left to right) are shown as dashed lines.
\label{fig1}}
\end{figure}

\clearpage

\begin{figure}
\includegraphics[scale=.65,angle=-90]{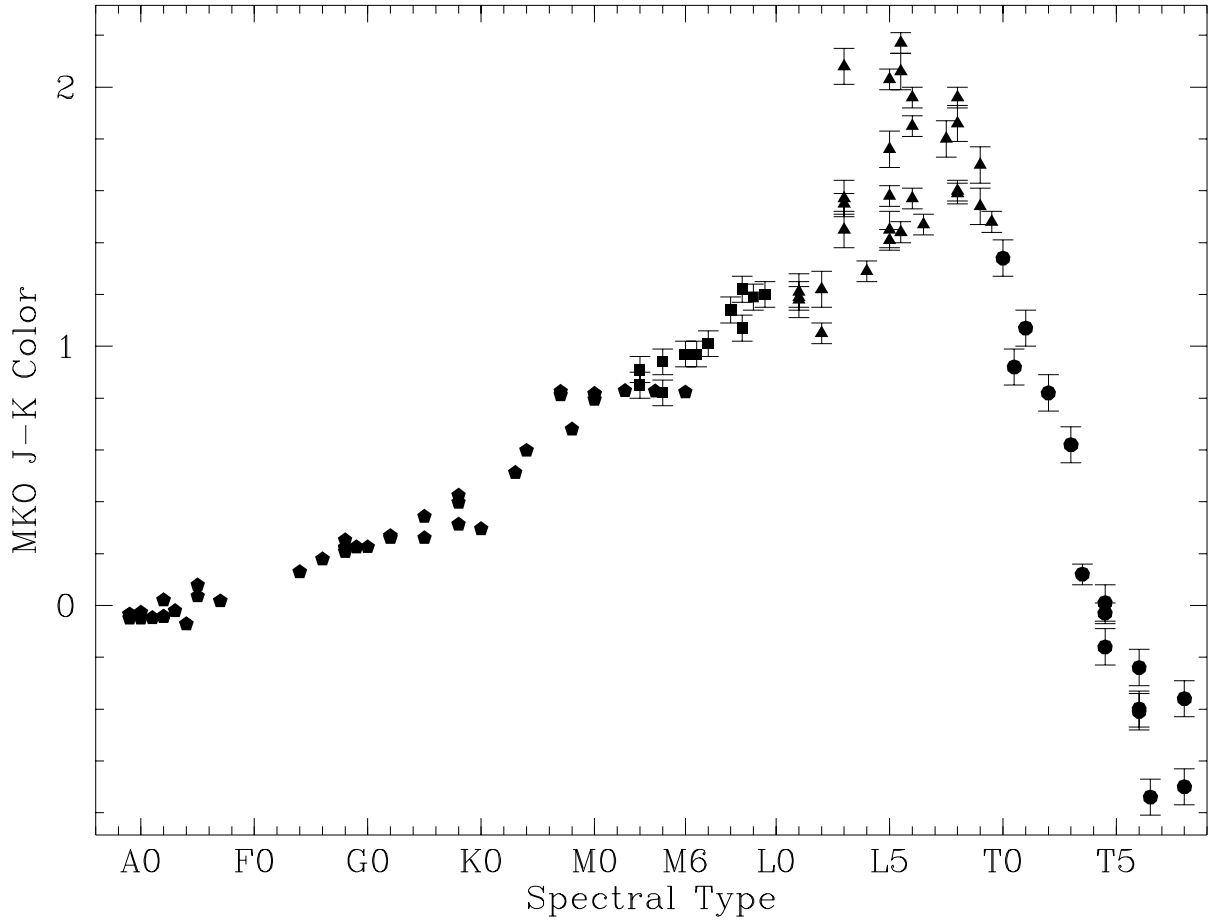}
\caption{Observed MKO $J-K$ colors for several late M (squares), L (triangles) and T dwarfs (circles) 
as a function of spectral type \citep{le02}.  Synthetic $J-K$ values generated for the standard main sequence 
stars in the Bruzual--Persson--Gunn--Stryker spectral atlas are also shown for comparison (pentagons).  
\label{fig2}}
\end{figure}

\clearpage
\begin{figure}
\includegraphics[scale=.65,angle=-90]{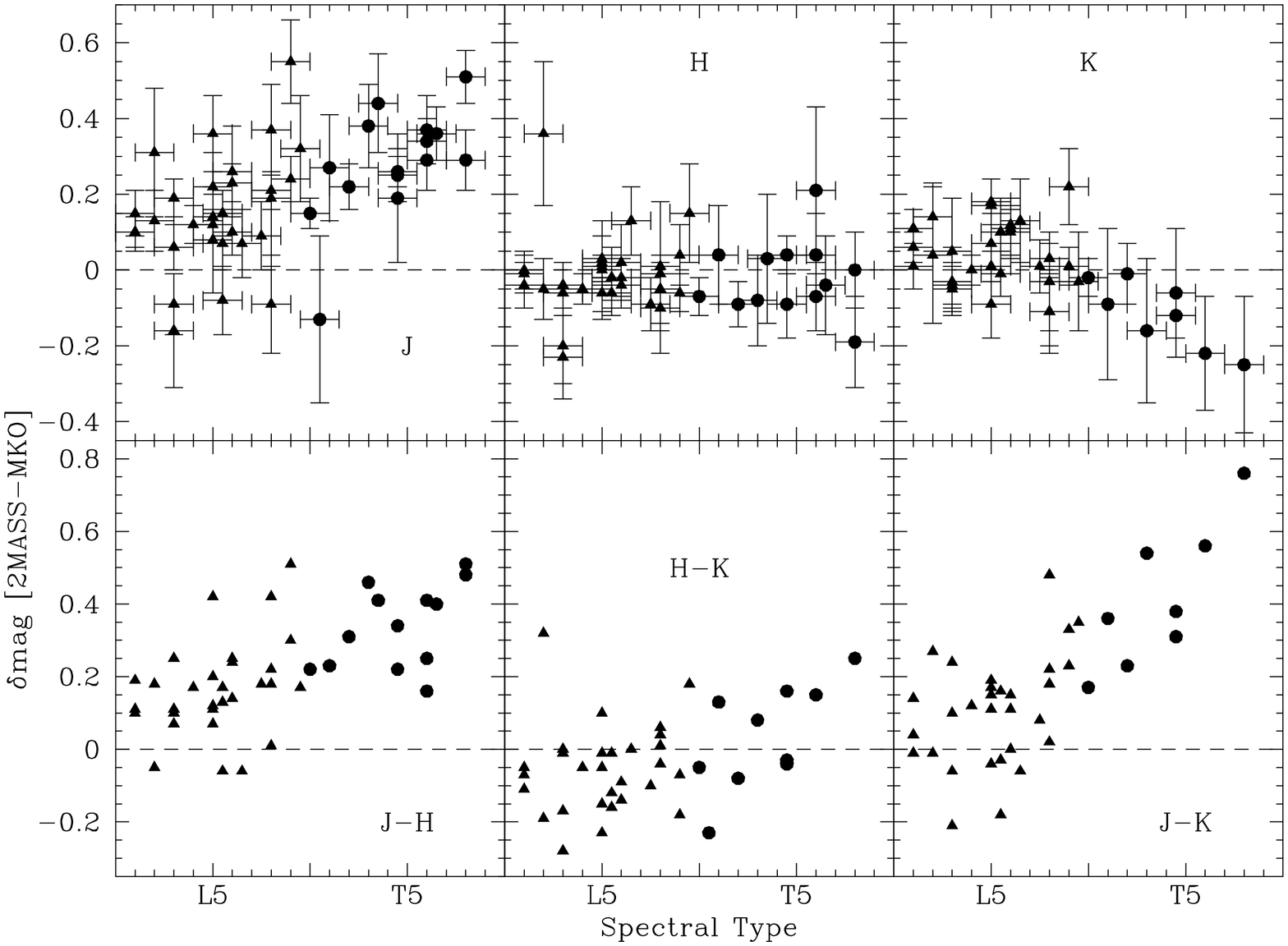}
\caption{Observed $\delta J$,  $\delta H$, $\delta K$, $\delta(J-H)$,  $\delta(H-K)$ and $\delta(J-K)$ mag,
as a function of spectral type, for the 2MASS and MKO systems.  
L dwarfs are shown as triangles and T dwarfs as circles. Error bars are omitted in the lower plot for clarity.
\label{fig3}}
\end{figure}

\clearpage

\begin{figure}
\includegraphics[scale=.65,angle=-90]{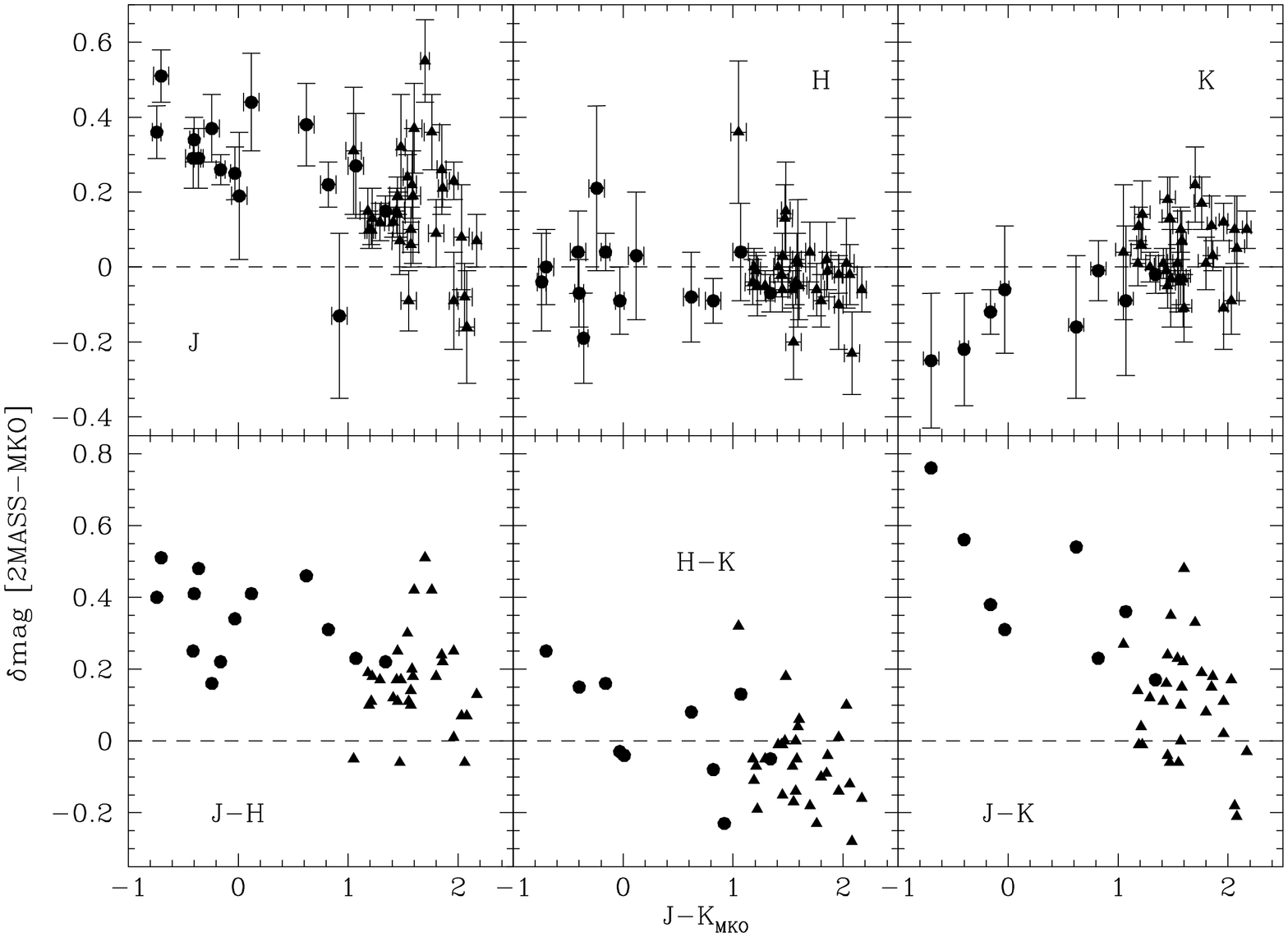}
\caption{Observed $\delta J$,  $\delta H$, $\delta K$, $\delta(J-H)$, $\delta(H-K)$ and $\delta(J-K)$ mag,
as a function of color, for the 2MASS and MKO systems.  L dwarfs are shown as triangles and T dwarfs
as circles.  Error bars are omitted in the lower plot for clarity.
\label{fig4}}
\end{figure}

\clearpage

\begin{figure}
\includegraphics[scale=0.80]{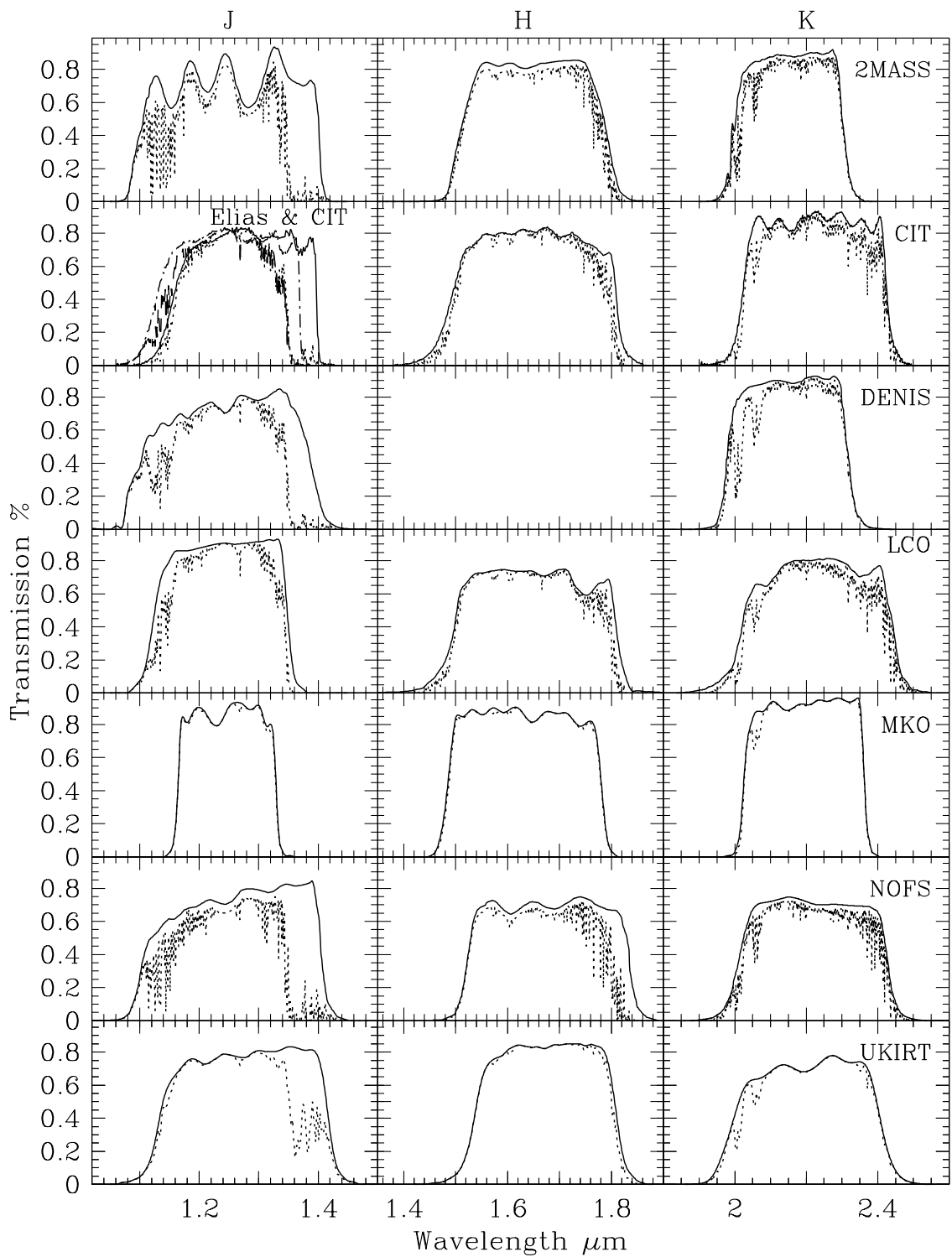}
\caption{Filter bandpasses for the systems considered here (solid line) and with
atmospheric absorption (dotted line).  The Elias--$J$ bandpass is drawn in the same 
box as the CIT--$J$ bandpass, both without atmospheric absorption (dash-dot line) and with (dashed line).
\label{fig5}}
\end{figure}

\clearpage

\begin{figure}
\includegraphics[scale=0.80]{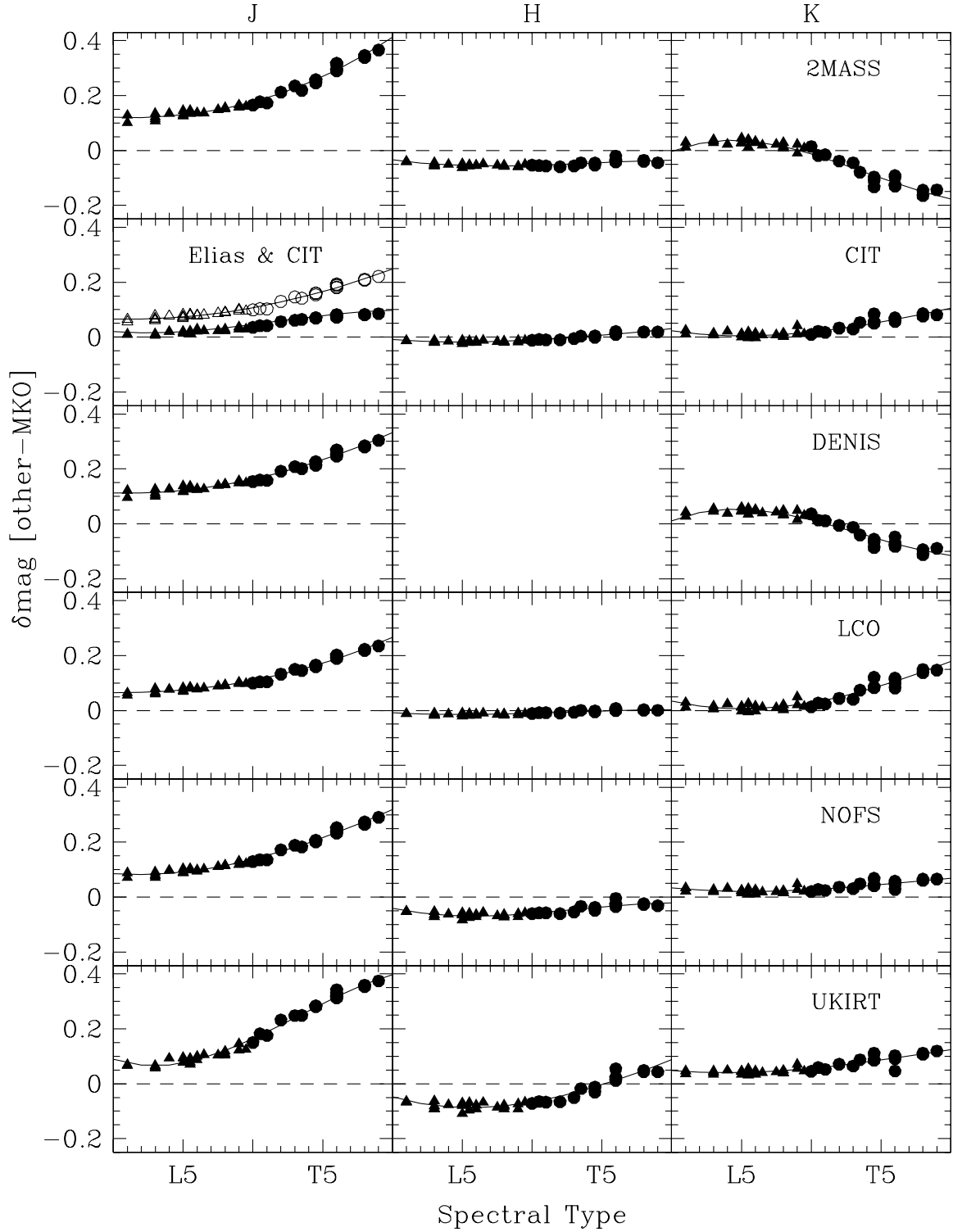}
\caption{Synthesized $\delta J$,  $\delta H$ and $\delta K$ mag,
as a function of spectral type for all the systems considered here.
L dwarfs are shown as triangles and T dwarfs
as circles.  Synthesized $\delta$mags using the Elias--$J$ filter are shown 
as open symbols.  The solid lines show the cubic fits given in Table 4.
\label{fig6}}
\end{figure}

\clearpage

\begin{figure}
\includegraphics[scale=0.80]{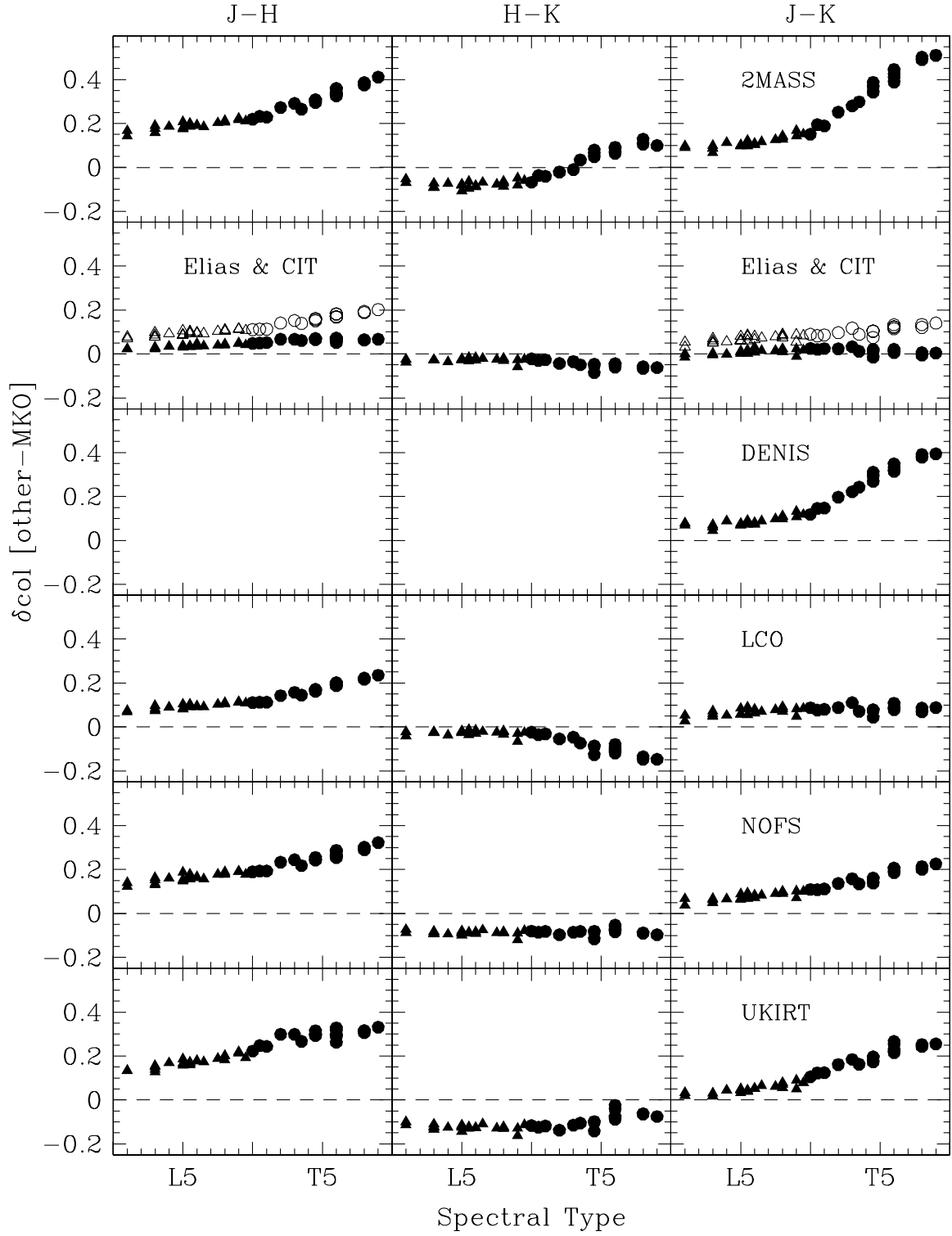}
\caption{Synthesized $\delta(J-H)$,  $\delta(H-K)$ and $\delta(J-K)$ mag,
as a function of spectral type for all the systems considered here.
Symbols are as in Figure 6.
\label{fig7}}
\end{figure}

\clearpage

\begin{figure}
\includegraphics[scale=0.80]{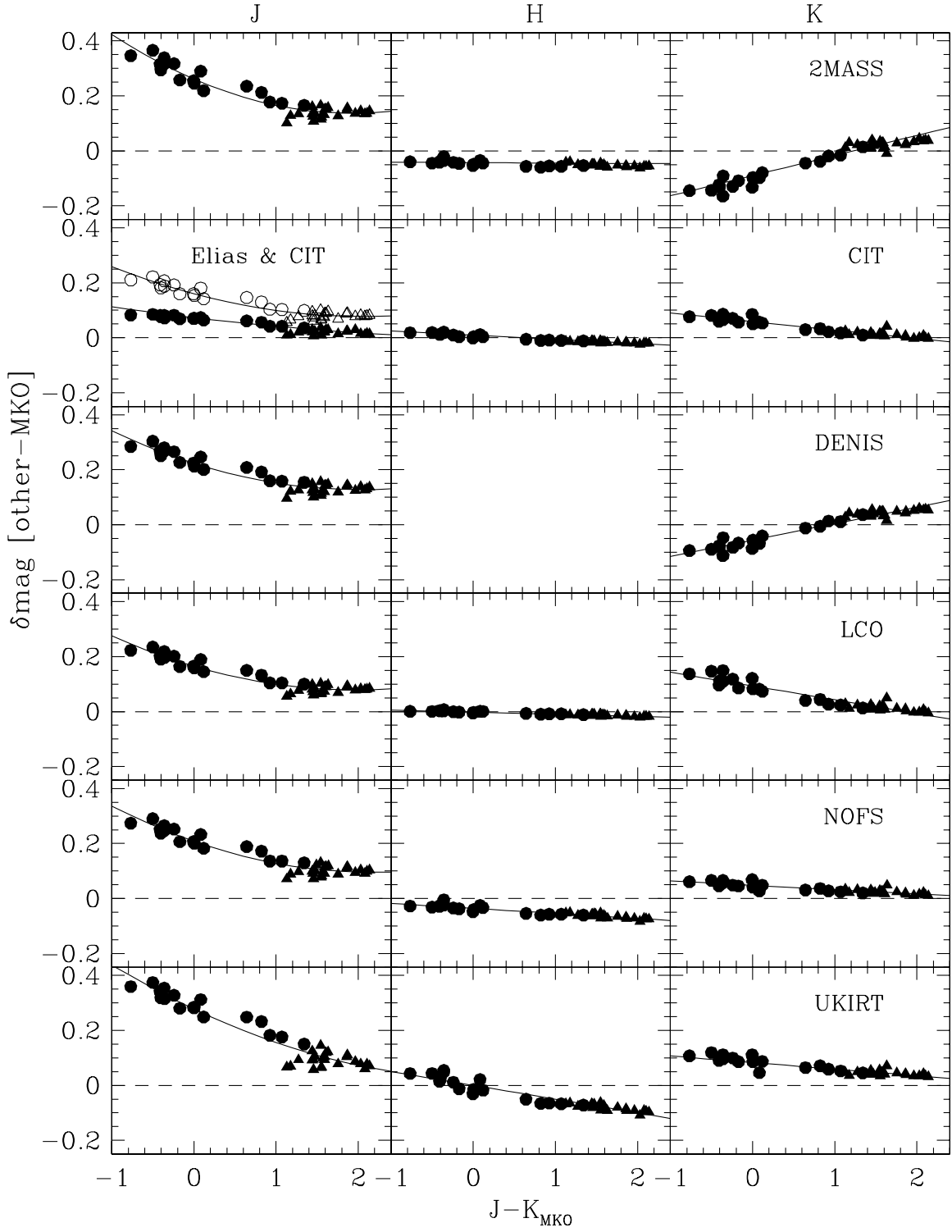}
\caption{Synthesized $\delta J$,  $\delta H$ and $\delta K$ mag,
as a function of color for all the systems considered here.  Symbols are as in Figure 6.
The solid lines show the fits given in Tables 5 and 6.
\label{fig8}}
\end{figure}

\clearpage

\begin{figure}
\includegraphics[scale=0.80]{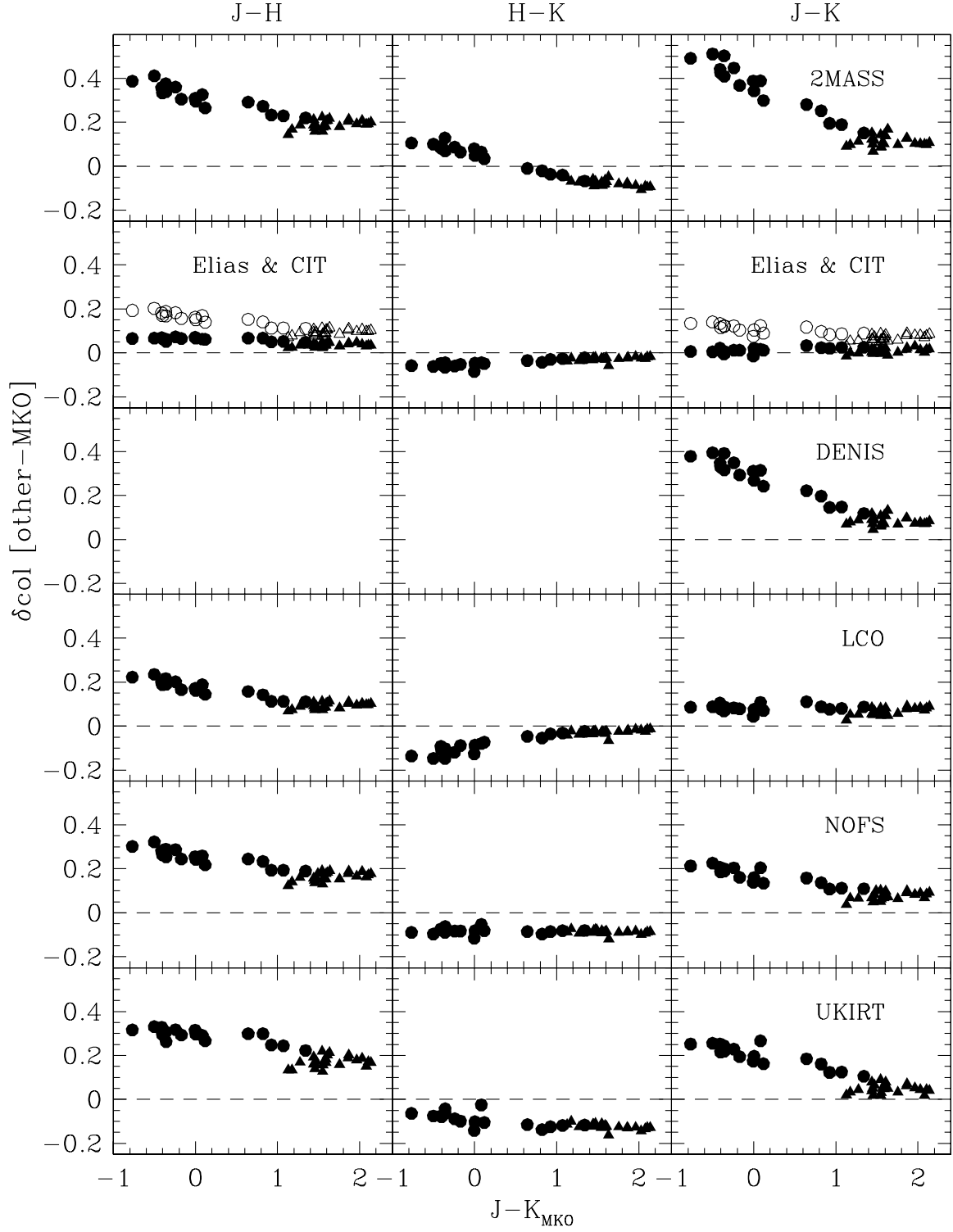}
\caption{Synthesized $\delta(J-H)$,  $\delta(H-K)$  and $\delta(J-K)$ mag,
as a function of color for all the systems considered here.  Symbols are as in Figure 6.
\label{fig9}}
\end{figure}

\clearpage

\begin{deluxetable}{llllllllll}
\tabletypesize{\scriptsize}
\tablewidth{0pt}
\tablecaption{Synthesized $J$-Band Photometry }
\tablehead{
\colhead{Name} & \colhead{Type} & \colhead{2MASS} &  \colhead{CIT} & \colhead{ELIAS} 
&  \colhead{DENIS} &  \colhead{LCO} &  \colhead{MKO} & \colhead{NOFS} & \colhead{UKIRT}
}
\tablecolumns{10}
\startdata 
2MASS0345$+$25  &  L1       & 13.967 & 13.854 & 13.904 & 13.961 & 13.903 & 13.839 & 13.928& 13.908 \\
2MASS0746$+$20AB  &  L1     & 11.683 & 11.595 & 11.637 & 11.677 & 11.637 & 11.581 &11.652 & 11.648 \\
2MASS0028$+$15  &  L3       & 16.788 & 16.668 & 16.731 & 16.779 & 16.733 & 16.653 & 16.744& 16.713 \\
DENIS1058$-$15  &  L3        & 14.244 & 14.136 & 14.188 & 14.234 & 14.188 & 14.121 &14.203 & 14.187 \\
GD165B        &  L3      & 15.752 & 15.653 & 15.702 & 15.743 & 15.702 & 15.637 & 15.715& 15.701  \\
Kelu$-$1      &  L3      & 13.343 & 13.246 & 13.295 & 13.335 & 13.295 & 13.235 &13.307 & 13.293  \\
2MASS0036$+$18  &  L4       & 12.454 & 12.344 & 12.396 & 12.445 & 12.396 & 12.319 &12.417 & 12.412 \\
SDSS0539$-$00  &  L5       & 13.976 & 13.876 & 13.923 & 13.967 & 13.922 & 13.850 &13.941 & 13.945  \\
SDSS1257$-$01  &  L5       & 15.770 & 15.663 & 15.713 & 15.759 & 15.712 & 15.639 & 15.731& 15.733 \\
SDSS1446$+$00  &  L5       & 15.686 & 15.577 & 15.628 & 15.677 & 15.628 & 15.559 & 15.647& 15.637  \\
SDSS2249$+$00  &  L5       & 16.608 & 16.482 & 16.544 & 16.600 & 16.545 & 16.462 &  16.566& 16.540 \\
DENIS0205$-$11AB  &  L5.5    & 14.540 & 14.431 & 14.485 & 14.531 & 14.486 & 14.405 & 14.504 & 14.497 \\
SDSS0107$+$00   &  L5.5    & 15.876 & 15.749 & 15.813 & 15.868 & 15.816 & 15.731 & 15.834& 15.802  \\
SDSS1326$-$00  &  L5.5     & 16.317 & 16.201 & 16.259 & 16.310 & 16.261 & 16.180 &  16.279& 16.258  \\
2MASS0825$+$21  &  L6       & 15.028 & 14.919 & 14.969 & 15.016 & 14.969 & 14.891 & 14.986& 14.978  \\
DENIS1228$-$15AB  &  L6      & 14.420 & 14.309 & 14.362 & 14.411 & 14.362 & 14.283 & 14.382& 14.383  \\
SDSS0236$+$00  &  L6.5     & 16.146 & 16.039 & 16.091 & 16.137 & 16.091 & 16.010 & 16.112& 16.116  \\
2MASS1632$+$19  &  L7.5     & 15.921 & 15.800 & 15.859 & 15.912 & 15.861 & 15.772 &15.883 & 15.876 \\
2MASS1523$+$30  &  L8       & 16.102 & 15.984 & 16.039 & 16.093 & 16.041 & 15.950 & 16.066& 16.069   \\
SDSS0032$+$14  &  L8       & 16.733 & 16.609 & 16.670 & 16.724 & 16.672 & 16.581 &16.695 & 16.686  \\
SDSS0857$+$57  &  L8       & 14.956 & 14.830 & 14.891 & 14.945 & 14.893 & 14.800 &14.916 & 14.910  \\
2MASS0310$+$16  &  L9       & 15.996 & 15.873 & 15.934 & 15.986 & 15.936 & 15.838 &15.957 & 15.960  \\
SDSS0830$+$48  &  L9       & 15.389 & 15.265 & 15.324 & 15.379 & 15.326 & 15.223 &15.354 & 15.369  \\
2MASS0328$+$23  &  L9.5     & 16.511 & 16.383 & 16.445 & 16.499 & 16.448 & 16.350 & 16.472& 16.475 \\
SDSS0423$-$04  &  T0       & 14.466 & 14.342 & 14.400 & 14.455 & 14.401 & 14.301 &14.430 & 14.451 \\
SDSS0837$-$00  &   T0.5    & 17.079 & 16.949 & 17.006 & 17.061 & 17.006 & 16.902 &17.038 & 17.084 \\
SDSS0151$+$12  &  T1       & 16.424 & 16.298 & 16.354 & 16.409 & 16.355 & 16.251 & 16.387 & 16.427  \\
SDSS1254$-$01  &   T2      & 14.873 & 14.724 & 14.791 & 14.852 & 14.793 & 14.661 &14.833 & 14.893  \\
SDSS1021$-$03  &  T3       & 16.115 & 15.948 & 16.026 & 16.088 & 16.030 & 15.880 &16.069 & 16.128  \\
SDSS1750$+$17  &  T3.5     & 16.358 & 16.210 & 16.281 & 16.340 & 16.284 & 16.139 &16.322 & 16.388  \\
2MASS0559$-$14  &  T4.5     & 13.829 & 13.648 & 13.731 & 13.797 & 13.735 & 13.571 &13.777 & 13.851  \\
SDSS0207$+$00  &  T4.5     & 16.886 & 16.709 & 16.791 & 16.855 & 16.796 & 16.631 & 16.838 & 16.915 \\
SDSS0926$+$58  &   T4.5    & 15.714 & 15.546 & 15.622 & 15.680 & 15.627 & 15.468 &15.669 & 15.750   \\
2MASS1225$-$27AB  &  T6     & 15.173 & 14.963 & 15.061 & 15.130 & 15.070 & 14.879 & 15.116 & 15.197  \\
GL229B  &  T6            & 14.324 & 14.087 & 14.195 & 14.275 & 14.204 & 14.007 & 14.255 & 14.323  \\
SDSS1110$+$01  &  T6       & 16.411 & 16.202 & 16.301 & 16.367 & 16.310 & 16.121 &  16.354& 16.433  \\
SDSS1346$-$00  &   T6      & 15.810 & 15.582 & 15.685 & 15.758 & 15.694 & 15.493 &  15.745 & 15.821 \\
SDSS1624$+$00  &  T6       & 15.513 & 15.287 & 15.390 & 15.466 & 15.398 & 15.197 & 15.449 & 15.539  \\
2MASS1217$-$03  &  T8       & 15.900 & 15.652 & 15.769 & 15.841 & 15.780 & 15.562 &15.827 & 15.916  \\
GL570D  &   T8           & 15.101 & 14.846 & 14.966 & 15.039 & 14.978 & 14.755 & 15.029 & 15.114 \\
2MASS0415$-$09  &  T9       & 15.687 & 15.414 & 15.542 & 15.625 & 15.556 & 15.321 & 15.611& 15.695  \\
\enddata

\end{deluxetable}

\clearpage

\begin{deluxetable}{llllllll}
\tabletypesize{\scriptsize}
\tablewidth{0pt}
\tablecaption{Synthesized $H$-Band Photometry }
\tablehead{
\colhead{Name} & \colhead{Type} & \colhead{2MASS} &  \colhead{CIT} 
&  \colhead{LCO} &  \colhead{MKO} & \colhead{NOFS} & \colhead{UKIRT}
}
\tablecolumns{8}
\startdata 
2MASS0345$+$25 & L1     & 13.169 & 13.195 & 13.197 & 13.208 &13.157 & 13.144  \\
2MASS0746$+$20AB & L1   & 10.943 & 10.970 & 10.972 & 10.984 & 10.931& 10.917  \\
2MASS0028$+$15 & L3     & 15.514 & 15.550 & 15.553 & 15.570 &15.498 & 15.479 \\
DENIS1058$-$15 & L3      & 13.232 & 13.266 & 13.269 & 13.286 &13.214 & 13.196  \\
GD165B       & L3    & 14.705 & 14.737 & 14.739 & 14.749 &14.696 & 14.686  \\
Kelu$-$1    & L3     & 12.402 & 12.435 & 12.437 & 12.452 &12.388 & 12.371  \\
2MASS0036$+$18 & L4     & 11.547 & 11.583 & 11.585 & 11.598 & 11.536 & 11.522  \\
SDSS0539$-$00 & L5     & 12.991 & 13.028 & 13.030 & 13.040 &12.983 & 12.970  \\
SDSS1257$-$01 & L5     & 14.642 & 14.679 & 14.681 & 14.693 &14.631 & 14.620  \\
SDSS1446$+$00 & L5     & 14.536 & 14.572 & 14.575 & 14.588 &14.523 & 14.508  \\
SDSS2249$+$00 & L5     & 15.366 & 15.403 & 15.407 & 15.428 & 15.345& 15.319  \\
DENIS0205$-$11AB & L5.5  & 13.552 & 13.592 & 13.593 & 13.604 &13.545 & 13.535  \\
SDSS0107$+$00  & L5.5  & 14.506 & 14.540 & 14.543 & 14.561 &14.486 & 14.464  \\
SDSS1326$-$00 & L5.5   & 14.968 & 15.004 & 15.007 & 15.024 &14.950 & 14.930   \\
2MASS0825$+$21 & L6     & 13.755 & 13.790 & 13.792 & 13.810 &13.738 & 13.718   \\
DENIS1228$-$15AB & L6    & 13.354 & 13.394 & 13.395 & 13.408 &13.344 & 13.330   \\
SDSS0236$+$00 & L6.5   & 15.112 & 15.149 & 15.151 & 15.161 & 15.105& 15.093  \\
2MASS1632$+$19 & L7.5   & 14.683 & 14.720 & 14.723 & 14.737 &14.669 & 14.652 \\
2MASS1523$+$30 & L8     & 15.008 & 15.048 & 15.050 & 15.064 & 14.997& 14.982   \\
SDSS0032$+$14 & L8     & 15.609 & 15.648 & 15.650 & 15.663 &15.599 & 15.585  \\
SDSS0857$+$57 & L8     & 13.755 & 13.794 & 13.796 & 13.813 & 13.740 & 13.721   \\
2MASS0310$+$16 & L9     & 14.852 & 14.893 & 14.895 & 14.911 &14.839 & 14.820  \\
SDSS0830$+$48 & L9     & 14.345 & 14.391 & 14.393 & 14.402 &14.341 & 14.329  \\
2MASS0328$+$23 & L9.5   & 15.432 & 15.471 & 15.472 & 15.483 & 15.426& 15.416  \\
SDSS0423$-$04 & T0     & 13.456 & 13.498 & 13.499 & 13.510 &13.449 & 13.438 \\
SDSS0837$-$00 &  T0.5  & 16.155 & 16.201 & 16.202 & 16.210 &16.153 & 16.145   \\	
SDSS0151$+$12 & T1     & 15.484 & 15.531 & 15.532 & 15.540 & 15.482& 15.473  \\
SDSS1254$-$01 &  T2    & 14.070 & 14.121 & 14.120 & 14.130 &14.069 & 14.064   \\
SDSS1021$-$03 & T3     & 15.376 & 15.428 & 15.425 & 15.432 &15.377 & 15.381   \\
SDSS1750$+$17 & T3.5   & 15.894 & 15.943 & 15.939 & 15.939 &15.905 & 15.921   \\
2MASS0559$-$14 & T4.5   & 13.595 & 13.646 & 13.639 & 13.641 &13.603 & 13.628  \\
SDSS0207$+$00 & T4.5   & 16.581 & 16.634 & 16.628 & 16.634 &  16.586& 16.603   \\
SDSS0926$+$58 &  T4.5  & 15.400 & 15.454 & 15.446 & 15.449 & 15.408& 15.434  \\
2MASS1225$-$27AB & T6   & 15.131 & 15.183 & 15.171 & 15.169 &15.143 & 15.192  \\
GL229B & T6          & 14.337 & 14.380 & 14.365 & 14.358 & 14.353& 14.412   \\
SDSS1110$+$01 & T6     & 16.187 & 16.234 & 16.223 & 16.222 &16.196 & 16.243  \\
SDSS1346$-$00 &  T6    & 15.797 & 15.851 & 15.838 & 15.839 &15.804 & 15.850   \\
SDSS1624$+$00 & T6     & 15.441 & 15.496 & 15.485 & 15.482 & 15.453 & 15.496  \\
2MASS1217$-$03 & T8     & 15.942 & 15.999 & 15.980 & 15.978 &  15.954& 16.026  \\
GL570D &  T8         & 15.240 & 15.300 & 15.281 & 15.280 & 15.252& 15.323   \\
2MASS0415$-$09 & T9     & 15.652 & 15.718 & 15.697 & 15.697 & 15.665& 15.740  \\
\enddata

\end{deluxetable}

\clearpage

\begin{deluxetable}{lllllllll}
\tabletypesize{\scriptsize}
\tablewidth{0pt}
\tablecaption{Synthesized $K$-Band Photometry }
\tablehead{
\colhead{Name} & \colhead{Type} & \colhead{2MASS} &  \colhead{CIT} 
&  \colhead{DENIS} &  \colhead{LCO} &  \colhead{MKO} & \colhead{NOFS} & \colhead{UKIRT}
}
\tablecolumns{9}
\startdata 
2MASS0345$+$25 & L1     & 12.693 & 12.672 & 12.706 & 12.674 & 12.663 & 12.685 & 12.699 \\
2MASS0746$+$20AB & L1   & 10.464 & 10.475 & 10.479	& 10.481 & 10.452 &10.486 & 10.499 \\
2MASS0028$+$15 & L3     & 14.609 & 14.579 & 14.627 & 14.580 & 14.573 & 14.595 & 14.616 \\
DENIS1058$-$15 & L3      & 12.603 & 12.573 & 12.615	& 12.574 & 12.568 & 12.588 & 12.603 \\
GD165B    & L3       & 14.116 & 14.101 & 14.134	& 14.105 & 14.088 &  14.116 & 14.135  \\
Kelu$-$1     & L3    & 11.820 & 11.788 & 11.835	& 11.789 & 11.780 &11.803 & 11.820 \\
2MASS0036$+$18 & L4     & 11.067 & 11.063 & 11.083	& 11.069 & 11.045 & 11.077 & 11.094 \\
SDSS0539$-$00 & L5     & 12.427 & 12.409 & 12.446	& 12.413 & 12.400 &12.423 & 12.445 \\
SDSS1257$-$01 & L5     & 14.081 & 14.061 & 14.100 & 14.065 & 14.051 &14.076 & 14.097 \\
SDSS1446$+$00 & L5     & 13.832 & 13.811 & 13.851 & 13.814 & 13.803 & 13.827 & 13.848 \\
SDSS2249$+$00 & L5     & 14.478 & 14.432 & 14.493	& 14.431 & 14.433 &14.448 & 14.467 \\
DENIS0205$-$11AB & L5.5  & 12.977 & 12.982 & 13.000 & 12.990 & 12.965 & 12.996& 13.019 \\
SDSS0107$+$00  & L5.5  & 13.630 & 13.587 & 13.645	& 13.587 & 13.592 & 13.602 & 13.623 \\
SDSS1326$-$00 & L5.5   & 14.116 & 14.074 & 14.132	& 14.074 & 14.076 &14.089 & 14.110  \\
2MASS0825$+$21 & L6     & 12.961 & 12.924 & 12.980 & 12.925 & 12.928 &12.939 & 12.963  \\
DENIS1228$-$15AB & L6    & 12.741 & 12.725 & 12.760	& 12.729 & 12.713 & 12.740 & 12.760  \\
SDSS0236$+$00 & L6.5   & 14.559 & 14.546 & 14.579 & 14.551 & 14.540 &14.559 & 14.581 \\
2MASS1632$+$19 & L7.5   & 13.935 & 13.921 & 13.955	& 13.923 & 13.913 &13.933 & 13.956 \\
2MASS1523$+$30 & L8     & 14.358 & 14.348 & 14.378	& 14.352 & 14.343 &14.360 & 14.384 	 \\
SDSS0032$+$14 & L8     & 14.999 & 15.003 & 15.022 & 15.010 & 14.990 &15.014 & 15.039  \\
SDSS0857$+$57 & L8     & 12.960 & 12.932 & 12.980 & 12.934 & 12.932 &12.947 & 12.971  \\
2MASS0310$+$16 & L9     & 14.192 & 14.240 & 14.217	& 14.252 & 14.202 &14.250 & 14.273  \\
SDSS0830$+$48 & L9     & 13.702 & 13.691 & 13.728	& 13.699 & 13.679 &13.707 & 13.735  \\
2MASS0328$+$23 & L9.5   & 14.869 & 14.870 & 14.891	& 14.874 & 14.860 &14.881 & 14.906 \\
SDSS0423$-$04 & T0     & 12.974 & 12.968 & 12.996 & 12.973 & 12.960 & 12.980 & 13.005 \\
SDSS0837$-$00 &  T0.5  & 15.960 & 15.997 & 15.991 & 16.005 & 15.978 & 16.006& 16.037  \\	
SDSS0151$+$12 & T1     & 15.164 & 15.195 & 15.191	& 15.204 & 15.180 & 15.204 & 15.232  \\
SDSS1254$-$01 &  T2    & 13.800 & 13.870 & 13.833 & 13.883 & 13.839 & 13.875 & 13.910  \\
SDSS1021$-$03 & T3     & 15.193 & 15.265 & 15.225	& 15.278 & 15.238 &15.269 & 15.302  \\
SDSS1750$+$17 & T3.5   & 15.942 & 16.073 & 15.980	& 16.095 & 16.021 &  16.069& 16.108  \\
2MASS0559$-$14 & T4.5   & 13.634 & 13.800 & 13.676	& 13.829 & 13.743 & 13.788 & 13.829  \\
SDSS0207$+$00 & T4.5   & 16.504 & 16.721 & 16.549	& 16.757 & 16.636 & 16.704& 16.747  \\
SDSS0926$+$58 &  T4.5  & 15.367 & 15.515 & 15.408 & 15.547 & 15.464 &15.505 & 15.550  \\
2MASS1225$-$27AB & T6   & 15.153 & 15.350 & 15.203	& 15.394 & 15.282 &15.334 & 15.384  \\
GL229B & T6          & 14.275 & 14.435 & 14.318	& 14.476 & 14.366 &14.424 & 14.463   \\
SDSS1110$+$01 & T6     & 15.941 & 16.097 & 15.970	& 16.120 & 16.039 & 16.067 & 16.085 \\
SDSS1346$-$00 &  T6    & 15.607 & 15.810 & 15.653 & 15.854 & 15.736 & 15.784  & 15.835  \\
SDSS1624$+$00 & T6     & 15.484 & 15.669 & 15.530 & 15.704 & 15.608 &15.653 & 15.698 \\
2MASS1217$-$03 & T8     & 15.760 & 16.016 & 15.812	& 16.073 & 15.924 &15.989 & 16.035 \\
GL570D &  T8         & 15.379 & 15.607 & 15.430	& 15.661 & 15.524 & 15.585 & 15.631  \\
2MASS0415$-$09 & T9     & 15.679 & 15.909 & 15.733 & 15.970 & 15.823 & 15.888& 15.942  \\
\enddata

\end{deluxetable}

\clearpage
\voffset=0truein
\hoffset=0truein
\begin{deluxetable}{lclllll}
\tabletypesize{\scriptsize}
\tablewidth{0pt}
\tablecaption{ Coefficients of Cubic Fit\tablenotemark{a} ~to [$\delta$$mag$, Spectral Type]  }
\tablehead{
\colhead{System} &  \colhead{Filter} &  \colhead{Coefficient$_0$}
 &  \colhead{Coefficient$_1$}&  \colhead{Coefficient$_2$}&  \colhead{Coefficient$_3$}
 &\colhead{Error\tablenotemark{b}}  
}
\tablecolumns{7}
\startdata 
2MASS & $J$ &  $+$0.121 & $-$1.64e$-$3 & $+$6.32e$-$4  & $+$9.01e$-$6 & 4.5e$-$3 \\ 
2MASS & $H$ & $-$0.034 & $-$6.88e$-$3 & $+$6.27e$-$4  & $-$1.43e$-$5 & 1.1e$-$3 \\
2MASS & $K$ & $-$0.004 & $+$2.04e$-$2 &  $-$2.80e$-$3  & $+$6.75e$-$5 & 7.8e$-$3 \\
CIT & $J$ & $+$0.020 & $-$3.97e$-$3 & $+$9.20e$-$4 & $-$2.64e$-$5 & 0.6e$-$3 \\
CIT & $H$ & $-$0.009 & $-$3.69e$-$3  &  $+$4.22e$-$4 & $-$6.92e$-$6 & 0.5e$-$3 \\
CIT & $K$ & $+$0.023 & $-$7.39e$-$3 & $+$8.11e$-$4 &  $-$1.17e$-$5 & 2.3e$-$3 \\
ELIAS & $J$ & $+$0.066 & $-$1.16e$-$3 & $+$5.45e$-$4 & $-$1.57e$-$6 & 1.4e$-$3 \\
DENIS & $J$ & $+$0.112 &  $-$1.16e$-$3 & $+$6.33e$-$4 & $-$1.21e$-$6 & 3.2e$-$3 \\
DENIS & $K$ & $+$0.011 & $+$1.99e$-$2 & $-$2.60e$-$3  &  $+$6.44e$-$5 & 5.8e$-$3 \\
LCO & $J$ &  $+$0.065 & $-$4.82e$-$4 & $+$4.52e$-$4 & $+$3.74e$-$6 & 1.7e$-$3 \\
LCO & $H$ & $-$0.008 & $-$3.11e$-$3 & $+$4.00e$-$4 & $-$1.09e$-$5 & 0.3e$-$3 \\
LCO & $K$ & $+$0.034 & $-$1.13e$-$2 &  $+$1.22e$-$3 & $-$1.47e$-$5 & 6.3e$-$3 \\
NOFS & $J$ & $+$0.085 & $-$3.27e$-$3 & $+$9.62e$-$4 & $-$1.07e$-$5 & 2.5e$-$3 \\
NOFS & $H$ & $-$0.041 &  $-$1.08e$-$2 &  $+$1.21e$-$3 &  $-$3.11e$-$5 & 2.5e$-$3 \\
NOFS & $K$ & $+$0.034 &  $-$5.27e$-$3 & $+$5.65e$-$4 & $-$1.07e$-$5 & 2.8e$-$3 \\
UKIRT & $J$ & $+$0.089 & $-$1.67e$-$2 & $+$3.23e$-$3 & $-$8.11e$-$5 & 5.3e$-$3 \\
UKIRT & $H$ & $-$0.048 & $-$1.44e$-$2 & $+$1.45e$-$3 & $-$1.96e$-$5 & 6.7e$-$3 \\
UKIRT & $K$ & $+$0.048 & $-$4.15e$-$3 & $+$5.96e$-$4 & $-$9.94e$-$6 & 5.0e$-$3 \\
\enddata
\tablenotetext{a}{Fit is applied as 
$$mag_{system} - mag_{MKO} = Coeff_{0} + Coeff_{1}\times Type +
Coeff_{2}\times Type^2  + Coeff_{3}\times Type^3 $$
where type is an integer such that 01$=$L1, 10$=$T0, 19$=$T9}
\tablenotetext{b}{RMS scatter of the fit in magnitudes}
\end{deluxetable}

\clearpage
\voffset=0truein
\hoffset=0truein
\begin{deluxetable}{lcllllc}
\tabletypesize{\scriptsize}
\tablewidth{0pt}
\tablecaption{Coefficients of Quadratic Fit\tablenotemark{a} ~to [$\delta$$J mag$, $J-K$]  }
\tablehead{
\colhead{System} & \colhead{Filter} & \colhead{Coefficient$_0$}
 &  \colhead{Coefficient$_1$} &  \colhead{Coefficient$_2$} & \colhead{Error\tablenotemark{b}}  
 &  \colhead{$J-K_{system}$}
}
\tablecolumns{7}
\startdata 
2MASS & J & $+$0.261 & $-$1.29e$-$01 & $+$3.35e$-$02 & $+$2.01e$-$02 & MKO
\\
CIT & J & $+$0.073 & $-$3.55e$-$02 & $+$4.09e$-$03 & $+$3.58e$-$03 & MKO
\\
ELIAS & J & $+$0.160 & $-$8.01e$-$02 & $+$1.95e$-$02 & $+$8.85e$-$03 & MKO
\\
DENIS & J & $+$0.223 & $-$9.48e$-$02 & $+$2.35e$-$02 & $+$1.37e$-$02 & MKO
\\
LCO & J & $+$0.165 & $-$8.79e$-$02 & $+$2.23e$-$02 & $+$1.02e$-$02 & MKO
\\
NOFS & J & $+$0.208 & $-$1.04e$-$01 & $+$2.40e$-$02 & $+$1.60e$-$02 & MKO
\\
UKIRT & J & $+$0.278 & $-$1.41e$-$01 & $+$1.93e$-$02 & $+$3.77e$-$02 & MKO
\\
\cutinhead{$J-K_{other}$}
2MASS & J & $+$0.325 & $-$1.93e$-$01 & $+$4.97e$-$02 & $+$2.54e$-$02 & 2MASS
\\
CIT & J & $+$0.074 & $-$3.56e$-$02 & $+$4.20e$-$03 & $+$3.74e$-$03 & CIT
\\
ELIAS & J & $+$0.168 & $-$8.74e$-$02 & $+$2.11e$-$02 & $+$9.43e$-$03 & ELIAS\\
DENIS & J & $+$0.257 & $-$1.29e$-$01 & $+$3.26e$-$02 & $+$1.57e$-$02 & DENIS
\\
LCO & J & $+$0.172 & $-$9.26e$-$02 & $+$2.31e$-$02 & $+$1.08e$-$02 & LCO
\\
NOFS & J & $+$0.227 & $-$1.22e$-$01 & $+$2.84e$-$02 & $+$1.79e$-$02 & NOFS
\\
UKIRT & J & $+$0.308 & $-$1.68e$-$01 & $+$2.61e$-$02 & $+$4.58e$-$02 & UKIRT
\\
\enddata
\tablenotetext{a}{Fit is applied as 
$$mag_{system} - mag_{MKO} = Coeff_{0} + Coeff_{1}\times J-K_{system} +
Coeff_{2}\times J-K_{system}^2  $$
where $J-K_{system}$ is the observed color in the photometric system of the 7th column}
\tablenotetext{b}{RMS scatter of the fit in magnitudes}
\end{deluxetable}

\clearpage
\voffset=0truein
\hoffset=0truein
\begin{deluxetable}{lcllllc}
\tabletypesize{\scriptsize}
\tablewidth{0pt}
\tablecaption{Coefficients of Linear Fit\tablenotemark{a} ~to [$\delta H$,$\delta K mag$, $J-K$]  }
\tablehead{
\colhead{System} & \colhead{Filter} &  \colhead{intercept} &  \colhead{err$_{int}$} 
&  \colhead{gradient} &  \colhead{err$_{grad}$} & \colhead{$J-K_{system}$} }  
\tablecolumns{7}
\startdata
2MASS & H  & $-$0.043 & $+$0.001 & $-$6.49e$-$03 & $+$1.07e$-$03 & MKO
\\
2MASS & K & $-$0.096 & $+$0.004 & $+$7.31e$-$02 & $+$3.23e$-$03 & MKO
\\
CIT & H & $+$0.009 & $+$0.001 & $-$1.52e$-$02 & $+$6.27e$-$04 & MKO
\\
CIT & K & $+$0.060 & $+$0.002 & $-$3.13e$-$02 & $+$1.70e$-$03 & MKO
\\
DENIS & K & $-$0.056 & $+$0.004 & $+$5.99e$-$02 & $+$2.66e$-$03 & MKO
\\
LCO & H & $-$0.002 & $+$0.001 & $-$7.46e$-$03 & $+$4.17e$-$04 & MKO
\\
LCO & K & $+$0.093 & $+$0.004  & $-$5.00e$-$02 & $+$2.66e$-$03 & MKO
\\
NOFS & H & $-$0.036 & $+$0.002 & $-$1.83e$-$02 & $+$1.21e$-$03 & MKO
\\
NOFS & K & $+$0.048 & $+$0.002 & $-$1.56e$-$02 & $+$1.40e$-$03 & MKO
\\
UKIRT  & H & $+$0.000 & $+$0.003 & $-$5.00e$-$02 & $+$2.50e$-$03 & MKO
\\
UKIRT  & K & $+$0.083 & $+$0.004 & $-$2.40e$-$02 & $+$3.04e$-$03 & MKO
\\
\cutinhead{$J-K_{other}$}
2MASS & H  & $-$0.040 & $+$0.002 & $-$7.77e$-$03 & $+$1.25e$-$03 & 2MASS
\\
2MASS & K & $-$0.126 & $+$0.007 & $+$8.34e$-$02 & $+$4.62e$-$03 & 2MASS
\\
CIT & H & $+$0.009 & $+$0.001 & $-$1.52e$-$02 & $+$6.25e$-$04 & CIT
\\
CIT & K & $+$0.060  & $+$0.002 & $-$3.13e$-$02 & $+$1.66e$-$03 & CIT
\\
DENIS & K & $-$0.074 & $+$0.005 & $+$6.74e$-$02 & $+$3.44e$-$03 & DENIS \\
LCO & H & $-$0.001 & $+$0.001 & $-$7.49e$-$03 & $+$4.24e$-$04 & LCO
\\
LCO & K & $+$0.097 & $+$0.004 & $-$5.03e$-$02 & $+$2.65e$-$03 & LCO
\\
NOFS & H & $-$0.032 & $+$0.002 & $-$1.94e$-$02 & $+$1.27e$-$03 & NOFS
\\
NOFS & K & $+$0.051 & $+$0.002 & $-$1.66e$-$02 & $+$1.46e$-$03 & NOFS
\\
UKIRT  & H & $+$0.009 & $+$0.004 & $-$5.41e$-$02 & $+$2.82e$-$03 & UKIRT \\
UKIRT  & K & $+$0.087 & $+$0.005 & $-$2.54e$-$02 & $+$3.42e$-$03 & UKIRT 
\\
\enddata
\tablenotetext{a}{Fit is applied as 
$$mag_{system} - mag_{MKO} = intercept \pm err_{int} + gradient \pm err_{grad}\times 
J-K_{system} $$ 
where $J-K_{system}$ is the observed color in the photometric system of the 7th column}
\end{deluxetable}

\end{document}